\documentclass[journal]{IEEEtran}
\IEEEoverridecommandlockouts

\usepackage[T1]{fontenc}
\usepackage{cite}
\usepackage{amsmath,amssymb,bm}
\usepackage{algorithm}
\usepackage{algpseudocode}
\usepackage{booktabs}
\usepackage{multirow}
\usepackage{graphicx}
\usepackage{xcolor}

\DeclareMathOperator{\Tr}{Tr}
\DeclareMathOperator{\diag}{diag}
\DeclareMathOperator{\Diag}{Diag}

\DeclareMathOperator{\vecop}{vec}
\DeclareMathOperator{\col}{col}
\newcommand{\C}{\mathbb{C}}
\newcommand{\R}{\mathbb{R}}
\newcommand{\E}{\mathbb{E}}
\newcommand{\CN}{\mathcal{CN}}
\newcommand{\jj}{\mathrm{j}}
\newcommand{\T}{^{\mathrm{T}}}
\newcommand{\herm}{^{\mathrm{H}}}
\newcommand{\mbf}[1]{\mathbf{#1}}
\newcommand{\norm}[1]{\left\lVert #1\right\rVert}
\newcommand{\Real}{\operatorname{Re}}

\usepackage{acro}
\DeclareAcronym{isac}{
  short = ISAC,
  long  = integrated sensing and communications
}
\DeclareAcronym{rf}{
  short = RF,
  long  = radio frequency
}
\DeclareAcronym{bs}{
  short = BS,
  long  = base station
}
\DeclareAcronym{genai}{
  short = GenAI,
  long  = generative AI
}
\DeclareAcronym{lidar}{
  short = LiDAR,
  long  = light detection and ranging
}
\DeclareAcronym{6g}{
  short = 6G,
  long  = sixth generation
}
\DeclareAcronym{5g}{
  short = 5G,
  long  = Fifth generation
}

\DeclareAcronym{ai}{
  short = AI,
  long  = artificial intelligence
}
\DeclareAcronym{3gpp}{
  short = 3GPP,
  long  = 3rd generation partnership project
}
\DeclareAcronym{slam}{
  short = SLAM,
  long  = simultaneous localization and mapping
}
\DeclareAcronym{csi}{
  short = CSI,
  long  = channel state information
}
\DeclareAcronym{ri}{
  short = R\&I,
  long  = research and innovation
}

\DeclareAcronym{jepa}{
  short = JEPA,
  long  = joint-embedding predictive architecture
}

\DeclareAcronym{mimo}{
  short = MIMO,
  long  = multiple-input multiple-output
}

\DeclareAcronym{ofdm}{
  short = OFDM,
  long  = orthogonal frequency-division multiplexing
}

\DeclareAcronym{ula}{
  short = ULA,
  long  = uniform linear array
}

\DeclareAcronym{cp}{
  short = CP,
  long  = cyclic prefix
}

\DeclareAcronym{rcs}{
  short = RCS,
  long  = radar cross-section
}

\DeclareAcronym{milp}{
  short = MILP,
  long  = mixed-integer linear programming
}

\DeclareAcronym{mi}{
  short = MI,
  long  = mixed-integer
}

\DeclareAcronym{lp}{
  short = LP,
  long  = linear programming
}

\DeclareAcronym{nr}{
  short = NR,
  long  = new radio
}

\DeclareAcronym{isl}{
  short = ISL,
  long  = integrated sidelobe level
}

\DeclareAcronym{snr}{
  short = SNR,
  long  = signal-to-noise ratio
}

\DeclareAcronym{cdf}{
  short = CDF,
  long  = cumulative distribution function
}

\DeclareAcronym{dd}{
  short = DD,
  long  = delay-Doppler
}

\DeclareAcronym{sca}{
  short = SCA,
  long  = successive convex approximation
}

\DeclareAcronym{socp}{
  short = SOCP,
  long  = second-order conic program
}

\DeclareAcronym{dft}{
  short = DFT,
  long  = Discrete Fourier Transform
}

\DeclareAcronym{rhs}{
  short = RHS,
  long  = reconfigurable holographic surface
}

\DeclareAcronym{rd}{
  short = RD,
  long  = range-Doppler
}

\DeclareAcronym{fd}{
  short = FD,
  long  = fully digital
}

\DeclareAcronym{qcqp}{
  short = QCQP,
  long  = quadratically constrained quadratic program
}

\DeclareAcronym{dof}{
  short = DoF,
  long  = degrees of freedom
}

\DeclareAcronym{sinr}{
  short = SINR,
  long  = signal-to-interference-plus-noise ratio
}

\DeclareAcronym{zf}{
  short = ZF,
  long  = zero forcing
}

\DeclareAcronym{cfar}{
  short = CFAR,
  long  = constant false-alarm rate
}

\DeclareAcronym{csr}{
  short = CSR,
  long  = clutter-to-signal ratio
}

\begin{document}

\title{Holographic Beamforming for Range-Doppler Sidelobe Suppression in OFDM-ISAC}

\author{Amirhossein~Azarbahram~\IEEEmembership{Member,~IEEE} and Onel~L.~A.~L\'opez~\IEEEmembership{Senior~Member,~IEEE}\vspace{-6mm}
\\[1mm]
\thanks{Authors are with Centre for Wireless Communications (CWC), University of Oulu, Finland, (e-mail: \{amirhossein.azarbahram, onel.alcarazlopez\}@oulu.fi). This work is supported in Finland by the Research Council of Finland (Grant 369116 (6G Flagship)).}}

\maketitle

\begin{abstract}
This paper investigates range--Doppler (RD) sidelobe suppression in an integrated sensing and communications system with a reconfigurable holographic surface (RHS). We jointly design the digital feed precoders and RHS amplitudes to minimize the expected integrated RD sidelobe level over random communication symbols, subject to target-illumination and communication constraints. We then develop an alternating successive convex approximation method that updates the feed and RHS variables through convex quadratic subproblems. Numerical results show that the proposed design consistently reduces normalized RD sidelobe levels across different apertures and configurations, and improves sensing weak-target detectability.
\end{abstract}


\vspace{-4mm}
\section{Introduction}
\vspace{-1mm}

\IEEEPARstart{W}{ireless} systems are embracing \ac{isac} with communication infrastructure supporting radar functions. For this, \ac{ofdm} is attractive because its time-frequency structure supports data transmission while enabling \ac{rd} processing. The finite \ac{ofdm} grid generally produces \ac{rd} sidelobes masking weak targets. In \ac{mimo} systems, beamforming provides spatial \ac{dof}, constrained by transmit architecture, for shaping the \ac{rd} response~\cite{sturm2011waveform}.

\Acp{rhs} attracted interest as scalable apertures for wireless systems due to dense arrangement of subwavelength elements. \acp{rhs} enable wavefront shaping by controlling the radiation amplitudes of elements excited by a limited number of feeds, reducing hardware complexity. However, their feed-to-aperture mapping couples the digital feed with the surface amplitudes, preventing direct application of {fully digital} \ac{ofdm}-\ac{isac} precoding~\cite{zhang2022holographic_isac}.

Most \ac{isac} beamforming designs control the spatial beampattern, e.g.,~\cite{opt_BF_ISAC}. However, suppressing angular sidelobes alone does not suppress the \ac{rd} replicas produced by a finite multi-carrier block. Ambiguity-aware designs instead optimize \ac{ofdm} range sidelobes~\cite{LRS_DFRC}, \ac{mimo}-\ac{ofdm} \ac{rd} \ac{isl} through symbol-level waveform design~\cite{li2025mimo_ofdm_isl,li2025low_complexity}, local \ac{rd} sidelobes for orthogonal time frequency space~\cite{zhang2024otfs_wisl}, and range-angle sidelobes by block-level beamforming~\cite{liu2025range_angle}. Recently, movable antenna positions and beamforming were jointly optimized for \ac{rd} \ac{isl} suppression~\cite{feng2026ma_rd}. These works demonstrate the importance of hardware and spatial \ac{dof} for \ac{rd} sidelobe suppression, but do not consider \acp{rhs}.

\ac{rhs}-based \ac{isac} is studied, e.g., through joint digital-holographic beamforming~\cite{zhang2022holographic_isac}, sensing-information maximization~\cite{zhu2025rhs_isac}, and transmit-receive beamforming~\cite{yu2025joint_tx_rx_hisac}. Mutual-coupling-aware designs suppressed spatial sidelobes toward clutter directions~\cite{zeng2026mutual_coupling_isac} instead of shaping the \ac{rd} ambiguity surface. To the best of our knowledge, no work has investigated \ac{rd} sidelobe suppression for \ac{rhs}-aided \ac{ofdm} systems.

Here, we consider a multiuser \ac{mimo}-\ac{ofdm} \ac{isac} system employing an \ac{rhs} as the transmitter. Our contributions are threefold: i) we derive the expected \ac{rd} \ac{isl} in closed form, accounting for the symbol constellation through its second and fourth moment, and formulate the joint digital-feed/\ac{rhs} design under transmit-power, target-illumination, and per-user \ac{sinr} constraints; ii) we develop an alternating \ac{sca} solution with feasible inner approximations and analytical feed and amplitude gradients; and iii) we demonstrate numerically the benefits of the proposed design across different system dimensions and its improved weak-target detectability.

\textbf{\emph{Notation:}} Bold lowercase and uppercase letters denote vectors and matrices. $(\cdot)^T$, $(\cdot)^H$, and $(\cdot)^*$ denote transpose, Hermitian transpose, and complex conjugation; $\operatorname{Tr}(\cdot)$, $\operatorname{vec}(\cdot)$, $\operatorname{col}(\cdot)$, $\operatorname{Re}\{\cdot\}$, $\mathbb{E}[\cdot]$, and $\operatorname{Var}(\cdot)$ denote trace, vectorization, vertical concatenation, real part, expectation, and variance. $\operatorname{Diag}(\mathbf{x})$ forms a diagonal matrix from $\mathbf{x}$, while $\operatorname{diag}(\mathbf{X})$ extracts the main diagonal of $\mathbf{X}$. $\otimes$ and $\odot$ denote the Kronecker and Hadamard products, and $\|\cdot\|_2$ and $\|\cdot\|_F$ the Euclidean and Frobenius norms. $\mathbf{I}_N$, $\mathbf{0}$, $\mathbf{e}_k$, and $\mathbf{1}_{\{\cdot\}}$ denote the $N\times N$ identity matrix, an all-zero vector or matrix of appropriate size, the $k$-th canonical basis vector, and the indicator function.

\vspace{-4mm}
\section{System Model and Problem Formulation}

\begin{figure}
    \centering
    \includegraphics[width=0.85\columnwidth]{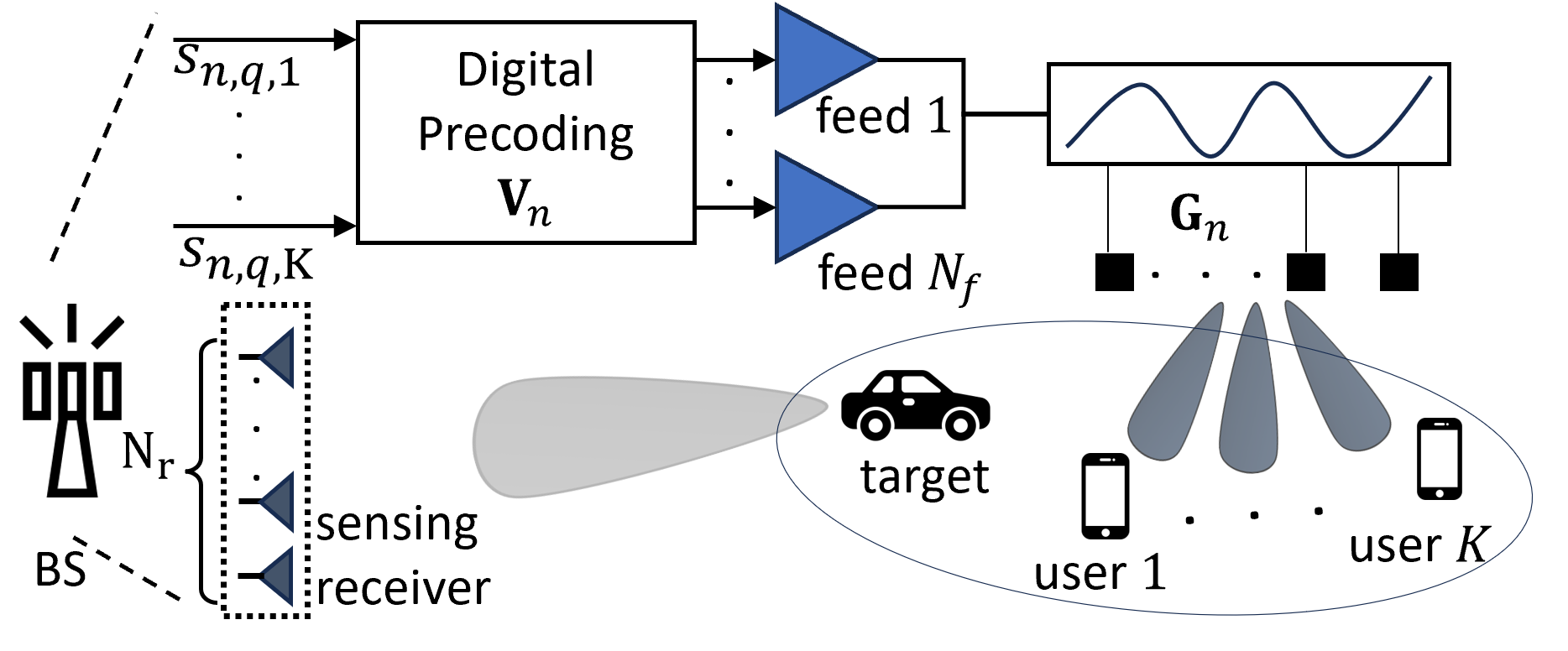}
   \caption{RHS-assisted monostatic \ac{ofdm}-\ac{isac} system.}
    \label{fig:sysmod}
    \vspace{-4mm}
\end{figure}

The system comprises a monostatic \ac{ofdm}-\ac{isac} \ac{bs} equipped with an \ac{rhs} consisting of $M$ elements and $N_{\rm f}$ feeds, with a co-located sensing receiver with $N_r$ antennas, as in \cite{zhang2022holographic_isac,zeng2026mutual_coupling_isac,zhu2025rhs_isac}. The \ac{bs} serves $K\leq N_{\rm f}$ single-antenna users and senses a target. The \ac{ofdm} block contains $N_c$ subcarriers and $N_s$ symbols. {The information-carrying vector on subcarrier $n$ and symbol $q$ is $\mbf{s}_{n,q}\in\mathbb{C}^{K}$, whose scalar entries are zero-mean, unit-average-power, independent and identically distributed, with $\rho\triangleq\E[s_{n,q,k}^2]$ and finite fourth moment $\mu_4\triangleq\E[|s_{n,q,k}|^4]<\infty$.\footnote{{Note that for most standard communication constellations, including normalized equiprobable QPSK and square QAM, properness holds ($\rho=0$).}}
} The digital precoder on subcarrier $n$ is $\mbf{V}_n=[\mbf{v}_{n,1},\ldots,\mbf{v}_{n,K}]\in\C^{N_{\rm f}\times K}$ and the feed-to-\ac{rhs} matrix is  $\mbf{G}_n\in\C^{M\times N_{\rm f}}$, whose $(m,r)$-th entry captures the propagation from feed $r$ to element $m$ on subcarrier $n$. The system model is illustrated in Fig.~\ref{fig:sysmod}.

The \ac{rhs} applies an amplitude-control vector
$\mbf{m}=[m_1,\ldots,m_M]\T\in[0,1]^M$, with
$\mbf{D}_{\mbf{m}}=\diag(\mbf{m})$. {This standard abstraction captures the dominant controllable \ac{rhs} response, while $\mbf G_n$ accounts for feed-to-element propagation. Element quantization, amplitude--phase coupling, dispersion, and mutual coupling are not explicitly represented, as in \cite{rhs_ampcontrol,zhu2025rhs_isac}.} The transmitted aperture-domain vector is given by
$\mbf{x}_{n,q}=\mbf{D}_{\mbf{m}}\mbf{G}_n\mbf{V}_n\mbf{s}_{n,q}
\in\C^M$, and the transmit power is
$
    P_{\rm tx}
    =\sum_{n=0}^{N_c-1}\Tr\!\left(
    \mbf{D}_{\mbf{m}}\mbf{G}_n\mbf{V}_n\mbf{V}_n\herm
    \mbf{G}_n\herm\mbf{D}_{\mbf{m}}
    \right).
    \label{eq:tx_power}
$

{
Let $\mbf{h}_{n,k}\in\C^M$ denote the channel from the \ac{rhs} elements to user $k$ on subcarrier $n$, which is assumed available and approximately constant over a block, as in~\cite{li2025mimo_ofdm_isl,opt_BF_ISAC,zhang2022holographic_isac}.} The received signal is written as
\begin{equation}
\vspace{-1mm}
    y_{n,q,k}=\mbf{h}_{n,k}\herm\mbf{D}_{\mbf{m}}\mbf{G}_n
    \mbf{V}_n\mbf{s}_{n,q}+z_{n,q,k},
    \label{eq:rx_comm}
    \vspace{-1mm}
\end{equation}
where $z_{n,q,k}\sim\CN(0,\sigma_c^2)$ is the noise. Then, the \ac{sinr} of user $k$ on subcarrier $n$ is given by
\begin{equation}
    {\rm \ac{sinr}}_{n,k}=
    \frac{|\mbf{h}_{n,k}\herm\mbf{D}_{\mbf{m}}\mbf{G}_n\mbf{v}_{n,k}|^2}
    {\sum_{j\neq k}|\mbf{h}_{n,k}\herm\mbf{D}_{\mbf{m}}\mbf{G}_n
    \mbf{v}_{n,j}|^2+\sigma_c^2}.
    \label{eq:sinr}
\end{equation}

\vspace{-4mm}
\subsection{Sensing Echo and Delay-Doppler Response}

We consider a point target at angle $\theta$, range $R$, and radial velocity $v$, with round-trip delay $\tau_0=2R/c_0$ and Doppler shift $f_d=2vf_c/c_0$. Let $\mbf a_{\rm T}(\theta)\in\C^M$ and $\mbf a_{\rm R}(\theta)\in\C^{N_r}$ denote the \ac{rhs} and receive steering vectors. For notational simplicity, we adopt one-dimensional azimuth propagation. The planar-\ac{rhs} extension follows by replacing the coordinates with their two-dimensional counterparts~\cite{zhang2022holographic_isac}. Assuming far-field narrowband propagation, synchronization and self-interference suppression, a block-static target, $\tau_0$ within the cyclic prefix, $|f_d|\ll\Delta f$ with negligible intercarrier interference, and $T_{\rm sym}$ as symbol duration, the echo on subcarrier $n$ and symbol $q$ is
\begin{equation}
    \mbf{y}^{\rm r}_{n,q}
    =\alpha_0\mbf{a}_{\rm R}(\theta)\mbf{a}_{\rm T}\herm(\theta)
    \mbf{x}_{n,q}
    e^{\jj2\pi(f_dqT_{\rm sym}-n\Delta f\tau_0)}
    +\mbf{n}^{\rm r}_{n,q},
    \label{eq:sensing_echo}
\end{equation}
where $\alpha_0$
captures the radar cross-section, propagation loss, and carrier phase, and
$\mbf{n}^{\rm r}_{n,q}\sim
\CN(\mbf{0},\sigma_r^2\mbf{I}_{N_r})$.

Unlike dedicated radar sequences, random communication symbols in \ac{isac} waveforms produce nonuniform directional powers and \ac{rd} sidelobes, which can be shaped through the feed precoders and \ac{rhs} amplitudes. Their aggregate energy is measured by the \ac{isl}. Define
$b_{n,q}=\mbf a_{\rm T}(\theta)\herm\mbf x_{n,q}$ and
$\mbf A(\theta)=\mbf a_{\rm T}(\theta)\mbf a_{\rm T}(\theta)\herm$.
After compensating for the nominal target delay and Doppler, the periodic \ac{rd} response is obtained by applying a two-dimensional discrete Fourier transform (DFT) to the directional resource-element powers
\cite{li2025mimo_ofdm_isl},
\begin{equation}
\begin{aligned}
\chi_{l,\nu}
&=\sum_{q=0}^{N_s-1}\sum_{n=0}^{N_c-1}
|b_{n,q}|^2 e^{-\jj2\pi ln/N_c}e^{\jj2\pi\nu q/N_s}
\\
&=\sum_{q=0}^{N_s-1}\sum_{n=0}^{N_c-1}
\left|\mbf a_{\rm T}\herm(\theta)\mbf D_{\mbf m}
\mbf G_n\mbf V_n\mbf s_{n,q}\right|^2
e^{-\jj2\pi ln/N_c}e^{\jj2\pi\nu q/N_s}.
\end{aligned}
\label{eq:ambiguity}
\end{equation}
Then, the \ac{isl} is given by $\Psi_{\rm ISL}=\sum_{(l,\nu)\in\Omega}|\chi_{l,\nu}|^2$, where $\mathcal{G}=\{0,\ldots,N_c-1\}\times\{0,\ldots,N_s-1\}, \Omega=\mathcal{G}\setminus\{(0,0)\}$.

\vspace{-4mm}
\subsection{Problem Formulation}

Reliable sensing requires both low \ac{rd} sidelobes and sufficient echo
strength. Since the received sensing \ac{snr} also depends on target
reflectivity and propagation loss, not directly controlled by the
\ac{bs}, we use the power radiated toward the target as a controllable proxy.
Assuming that the target direction $\theta$ is available from a preceding
detection stage, the symbol-averaged illumination power is
\begin{equation}
P_{\rm I}
=
\sum_{n=0}^{N_c-1}
\operatorname{Tr}\!\left(
\mbf V_n\herm
\mbf G_n\herm
\mbf D_{\mbf m}
\mbf A(\theta)
\mbf D_{\mbf m}
\mbf G_n
\mbf V_n
\right).
\label{eq:illumination_power}
\end{equation}
{Optimizing sidelobes for a particular symbol would require a fast
block-dependent redesign. We therefore aim to minimize
$\bar\Psi_{\rm ISL}\triangleq\E[\Psi_{\rm ISL}]$}, and write the optimization problem as
\begin{subequations}
\label{prob:main}
\begin{align}
    \min_{\{\mbf{V}_n\},\mbf{m}}\quad
    &{\bar\Psi_{\rm ISL}}
    \label{prob:main_obj}\\
    \text{s.t.}\quad
    &P_{\rm tx}\leq P_t,
    \label{prob:main_power}\\
    &P_{\rm I}\geq P_0,
    \label{prob:main_illum}\\
    &{\rm \ac{sinr}}_{n,k}\geq\gamma_{n,k},
    \qquad \forall n,k,
    \label{prob:main_sinr}\\
    &0\leq m_i\leq 1,
    \qquad i=1,\ldots,M.
    \label{prob:main_box}
\end{align}
\end{subequations}
Problem~\eqref{prob:main} is nonconvex because the variables are coupled, the illumination and
desired-signal terms are convex quadratic functions constrained from below,
and the objective is quartic in each variable block.{\footnote{Multiple sensing directions can be modeled by a weighted sum of the corresponding expected-\ac{isl} terms, while we focus on one sensing direction.}}

\vspace{-4mm}
\section{Optimization Framework}
\label{sec:sca_sca}


{We first derive the expected \ac{isl} by leveraging the symbol statistics. Define $c_{n,k}\triangleq
\mbf a_{\rm T}\herm(\theta)\mbf D_{\mbf m}
\mbf G_n\mbf v_{n,k}$,
$p_{n,k}\triangleq |c_{n,k}|^2$,
$a_n\triangleq\sum_k p_{n,k}$,
$t_n\triangleq\sum_k c_{n,k}^2$,
$z_{n,q}\triangleq|b_{n,q}|^2$, and $\vartheta_n\triangleq\operatorname{Var}(z_{n,q})$.
Under the assumed signaling, we can write
\begin{equation}
\begin{aligned}
\vartheta_n=a_n^2+(\mu_4-2)\sum_k p_{n,k}^2
+|\rho|^2\left(|t_n|^2-\sum_k p_{n,k}^2\right).
\end{aligned}
\label{eq:directional_moments}
\end{equation}
Using
$\E[z_{n,q}z_{n',q'}]
=a_na_{n'}+\mathbf 1_{\{n=n',q=q'\}}\vartheta_n$
in~\eqref{eq:ambiguity} yields
\begin{equation}
\E[|\chi_{l,\nu}|^2]
=\,N_s\sum_n\vartheta_n
+N_s^2\mathbf 1_{\{\nu=0\}}
\left|\sum_n a_ne^{-\jj2\pi ln/N_c}\right|^2.
\end{equation}
Summing over the \ac{rd} grid using DFT orthogonality
and excluding the main peak leads to
\begin{equation}
\bar\Psi_{\rm ISL}
=N_s(N_cN_s-1)\sum_n\vartheta_n
+N_s^2\left(N_c\sum_n a_n^2-P_{\rm I}^2\right),
\label{eq:expected_isl}
\end{equation}
where $P_{\rm I}=\sum_n a_n$.
After normalization by $N_s^2$, the objective depends on
$N_s$ only through $1/N_s$, while the design can be reused
across \ac{ofdm} blocks under an approximately unchanged channel,
target direction, and modulation statistics. $\bar\Psi_{\rm ISL}$ thus depends on the modulation statistics through both $\mu_4$ and $|\rho|^2$, including a phase-dependent coupling among user streams. Consequently, constellations with identical average power may exhibit different sidelobe energies depending on their higher-order statistics and beamforming design.
} We can then optimize the variables alternatively. Semidefinite relaxation can also apply, but it adds quadratic matrices and may yield high-rank solutions, infeasible to recover.

\vspace{-4mm}
\subsection{\ac{rhs}-Amplitude \ac{sca}}

{For both variable blocks, define
\begin{multline}
\xi_{n,k}
=
N_s(2N_cN_s-1)a_n-N_s^2P_{\rm I}
\\+N_s(N_cN_s-1)(\mu_4-2-|\rho|^2)p_{n,k}.
\label{eq:xi_def}
\end{multline}
}{Then, for fixed feed precoders, define
$\mbf U_n=\mbf G_n\mbf V_n$, $\mbf u_{n,k}=\mbf U_n\mbf e_k$, and
$\mbf H(\mbf c)\triangleq\Real\{\mbf c\mbf c\herm\}$.
With
$\mbf c_{n,k}^{\rm I}\triangleq
\mbf a_{\rm T}(\theta)^*\odot\mbf u_{n,k}$ and
$\mbf L_{n,k}\triangleq\mbf H(\mbf c_{n,k}^{\rm I})$,
we have $p_{n,k}=\mbf m\T\mbf L_{n,k}\mbf m$.
}
The transmit-power and illumination matrices are
\begin{equation}\label{eq:rhs_matrices}
\mbf P_{\rm m}(\mbf v)
\triangleq\Diag\!\left(\sum_n\diag(\mbf U_n\mbf U_n\herm)\right),
\
\mbf B_{\rm m}(\mbf v)
\triangleq\sum_{n,k}\mbf L_{n,k},
\end{equation}
such that
$\mbf m\T\mbf P_{\rm m}\mbf m=P_{\rm tx}$ and
$\mbf m\T\mbf B_{\rm m}\mbf m=P_{\rm I}$.
For communications, let
$
\mbf d_{n,k,j}\triangleq\mbf h_{n,k}^*\odot\mbf u_{n,j},
\mbf R_{n,k,j}\triangleq\mbf H(\mbf d_{n,k,j}),
$
and
$\mathcal J_{n,k}^{\rm m}
\triangleq\sum_{j\neq k}\mbf m\T\mbf R_{n,k,j}\mbf m+\sigma_c^2$.
{At iteration $r$,
\begin{multline}
\mbf g_{\rm m}^{(r)}
\!=\!{}
4\sum_{n,k}\xi_{n,k}^{(r)}
\mbf L_{n,k}\mbf m^{(r)}
\\
\!+\!
4N_s(N_cN_s-1)|\rho|^2
\Real\left\{
\sum_{n,k}(t_n^{(r)})^*
c_{n,k}^{(r)}\mbf c_{n,k}^{\rm I}
\right\}.
\label{eq:rhs_grad}
\end{multline}
\begin{multline}
\widehat f_{\rm m}^{(r)}(\mbf m)
={}
\bar\Psi_{\rm ISL}(\mbf v,\mbf m^{(r)})
\\
+
(\mbf g_{\rm m}^{(r)})\T
(\mbf m-\mbf m^{(r)})
+
\frac{\beta_{\rm m}^{(r)}}{2}
\norm{\mbf m-\mbf m^{(r)}}_2^2 .
\label{eq:rhs_surrogate}
\end{multline}}
The illumination and desired-signal lower bounds are
\begin{align}
\widetilde I_{\rm m}^{(r)}
&=
2(\mbf m^{(r)})\T\mbf B_{\rm m}\mbf m
-(\mbf m^{(r)})\T\mbf B_{\rm m}\mbf m^{(r)}, \label{eq:rhs_boundsa}
\\
\widetilde D_{n,k}^{\rm m,(r)}
&=
2(\mbf m^{(r)})\T\mbf R_{n,k,k}\mbf m
-(\mbf m^{(r)})\T\mbf R_{n,k,k}\mbf m^{(r)}.
\label{eq:rhs_boundsb}
\end{align}
Hence, the \ac{rhs} update is the convex \ac{qcqp} written as
\begin{subequations}\label{prob:rhs_sca}
\begin{align}
\min_{\mbf m\in\R^M}\quad
&\widehat f_{\rm m}^{(r)}(\mbf m)
\label{prob:rhs_sca_obj}\\
\text{s.t.}\quad
&\mbf m\T\mbf P_{\rm m}\mbf m\leq P_t,\quad
\widetilde I_{\rm m}^{(r)}\geq P_0,
\label{prob:rhs_sca_power}\\
&\widetilde D_{n,k}^{\rm m,(r)}
\geq\gamma_{n,k}\mathcal J_{n,k}^{\rm m},\ \forall n,k,\quad
0\leq m_i\leq1.
\label{prob:rhs_sca_sinr}
\end{align}
\end{subequations}

\vspace{-4mm}
\subsection{Feed-Domain \ac{sca}}

For fixed $\mbf m$, define
{$\mbf v=\col(\vecop(\mbf V_0),\ldots,\vecop(\mbf V_{N_c-1}))
\in\C^{D_v}$},
where $D_v=N_cN_{\rm f}K$.
Let $\mbf E_n\mbf v=\vecop(\mbf V_n)$ and {$\mbf E_{n,k}\mbf v=\mbf v_{n,k}$}.
Define
$\mbf T_n=\mbf G_n\herm\mbf D_{\mbf m}^2\mbf G_n$ and
$\mbf B_n=\mbf G_n\herm\mbf D_{\mbf m}
\mbf A(\theta)\mbf D_{\mbf m}\mbf G_n$, {Also, let
$\mbf w_n\triangleq
\mbf G_n\herm\mbf D_{\mbf m}\mbf a_{\rm T}(\theta)$,
such that
$c_{n,k}=\mbf w_n\herm\mbf E_{n,k}\mbf v$.
}
Then, we write
\begin{align}
\mbf P_{\rm v}
\!\triangleq\!\sum_n\mbf E_n\herm
(\mbf I_K\!\otimes\!\mbf T_n)\mbf E_n,\
\mbf B_{\rm v}
\!\triangleq\!\sum_n\mbf E_n\herm
(\mbf I_K\otimes\mbf B_n)\mbf E_n,
\end{align}
with
$\mbf v\herm\mbf P_{\rm v}\mbf v=P_{\rm tx}$ and
$\mbf v\herm\mbf B_{\rm v}\mbf v=P_{\rm I}$.
We then write
\begin{equation}
\mbf q_{n,k}\triangleq\mbf G_n\herm\mbf D_{\mbf m}\mbf h_{n,k},
\quad
\mbf C_{n,k,j}^{\rm v}
\triangleq
\mbf E_{n,j}\herm\mbf q_{n,k}\mbf q_{n,k}\herm\mbf E_{n,j},
\end{equation}
and
$\mathcal J_{n,k}^{\rm v}
\triangleq\sum_{j\neq k}\mbf v\herm
\mbf C_{n,k,j}^{\rm v}\mbf v+\sigma_c^2$.
{Since
$p_{n,k}=\mbf v\herm\mbf E_{n,k}\herm
\mbf B_n\mbf E_{n,k}\mbf v$, we can write
\begin{equation}
\begin{aligned}
\mbf g_{\rm v}^{(r)}
={}&
2\sum_{n,k}\xi_{n,k}^{(r)}
\mbf E_{n,k}\herm\mbf B_n\mbf E_{n,k}\mbf v^{(r)}
\\
&+
2N_s(N_cN_s-1)|\rho|^2
\sum_{n,k}
t_n^{(r)}(c_{n,k}^{(r)})^*
\mbf E_{n,k}\herm\mbf w_n.
\end{aligned}
\label{eq:feed_grad}
\end{equation}
The proximal model is given by
\begin{align}
\widehat f_{\rm v}^{(r)}(\mbf v)
=&\,\bar\Psi_{\rm ISL}(\mbf v^{(r)},\mbf m)
+2\Real\!\left\{(\mbf g_{\rm v}^{(r)})\herm
(\mbf v-\mbf v^{(r)})\right\}
\notag\\
&+\frac{\beta_{\rm v}^{(r)}}{2}
\norm{\mbf v-\mbf v^{(r)}}_2^2 .
\label{eq:feed_surrogate}
\end{align}
}

{The feed-domain lower bounds
$\widetilde I_{\rm v}^{(r)}$ and
$\widetilde D_{n,k}^{\rm v,(r)}$
follow from~\eqref{eq:rhs_boundsa} and \eqref{eq:rhs_boundsb} by replacing
$(\mbf m,\mbf m^{(r)},\mbf B_{\rm m},\mbf R_{n,k,k})$
with
$(\mbf v,\mbf v^{(r)},\mbf B_{\rm v},\mbf C_{n,k,k}^{\rm v})$
and using $2\Real\{\cdot\}$ for the complex bilinear terms.
}
Thus, the feed problem becomes
\begin{subequations}\label{prob:feed_sca}
\begin{align}
\min_{\mbf v}\quad
&\widehat f_{\rm v}^{(r)}(\mbf v)
\label{prob:feed_sca_obj}\\
\text{s.t.}\quad
&\mbf v\herm\mbf P_{\rm v}\mbf v\leq P_t,\quad
\widetilde I_{\rm v}^{(r)}\geq P_0,
\label{prob:feed_sca_power}\\
&\widetilde D_{n,k}^{\rm v,(r)}
\geq\gamma_{n,k}\mathcal J_{n,k}^{\rm v},
\qquad\forall n,k .
\label{prob:feed_sca_sinr}
\end{align}
\end{subequations}

\vspace{-4mm}
\subsection{Algorithm Description, Convergence, and Complexity}

{For $\mbf x\in\{\mbf v,\mbf m\}$, let
$f_{\rm x}(\mbf x;\mbf y)$ denote $\bar\Psi_{\rm ISL}$ with the other block
$\mbf y$ fixed, and let $\widehat f_{\rm x}^{(r)}$ denote the corresponding proximal model.
}
Algorithm~\ref{alg:sca_sca} starts from a feasible pair and alternately
updates the feed and \ac{rhs} {by invoking the
\textsc{BlockSCA} solver in Algorithm~\ref{alg:block_sca}}. For either
block $\mbf{x}\in\{\mbf{v},\mbf{m}\}$, the tight global lower bounds preserve
feasibility. {Within Algorithm~\ref{alg:block_sca},} backtracking
increases $\beta_{\rm x}$ by $\tau_\beta>1$ until satisfying the constraints and
$f_{\rm x}(\mbf{x}^{+};\mbf y)\leq
\widehat f_{\rm x}^{(r)}(\mbf{x}^{+};\mbf y)$. Consequently,
\begin{equation}
f_{\rm x}(\mbf{x}^{+};\mbf y)
\leq\widehat f_{\rm x}^{(r)}(\mbf{x}^{+};\mbf y)
\leq\widehat f_{\rm x}^{(r)}(\mbf{x}^{(r)};\mbf y)
=f_{\rm x}(\mbf{x}^{(r)};\mbf y),
\label{eq:descent_chain}
\end{equation}
so every accepted update yields a non-increasing
{expected \ac{isl}}. The inner and outer loops terminate according
to $\epsilon_{\rm in}$ and $\epsilon_{\rm out}$.

\begin{algorithm}[t]
\caption{\ac{sca}-based alternating \ac{rhs} optimization}
\label{alg:sca_sca}
\begin{algorithmic}[1]

\Require $\{\mbf G_n\}$, $\{\mbf h_{n,k}\}$, {$\mu_4$, $\rho$}, $\theta$,
$P_t$, $P_0$, $\{\gamma_{n,k}\}$,
$\beta_{\rm v},\beta_{\rm m}>0$, $I_{\max}$, $\epsilon_{\rm out}$, and feasible
$(\mbf v^{(0)},\mbf m^{(0)})$.

\State $(\mbf v_{\rm best},\mbf m_{\rm best})
\gets(\mbf v^{(0)},\mbf m^{(0)})$, $\zeta\gets\textsc{capped}$.

\For{$i=0,\ldots,I_{\max}-1$}

    \State $(\mbf v^{(i+1)},\zeta_{\rm v})
    \gets\textsc{BlockSCA}
    (\mbf v^{(i)},\mbf m^{(i)},\eqref{prob:feed_sca},\beta_{\rm v})$;
    evaluate $\bar\Psi_{\rm ISL}$ and update the best feasible pair.

    \If{$\zeta_{\rm v}=\textsc{stalled}$}
        $\zeta\gets\textsc{stalled}$; \textbf{break}
    \EndIf

    \State $(\mbf m^{(i+1)},\zeta_{\rm m})
    \gets\textsc{BlockSCA}
    (\mbf m^{(i)},\mbf v^{(i+1)},\eqref{prob:rhs_sca},\beta_{\rm m})$;
    evaluate $\bar\Psi_{\rm ISL}$ and update the best feasible pair.

    \If{$\zeta_{\rm m}=\textsc{stalled}$}
        $\zeta\gets\textsc{stalled}$; \textbf{break}
    \EndIf

    \If{outer convergence criterion is satisfied}
        \State $\zeta\gets\textsc{converged}$; \textbf{break}
    \EndIf

\EndFor

\State \Return best feasible
$(\{\mbf V_n^\star\},\mbf m^\star)$ and $\zeta$.

\end{algorithmic}

\end{algorithm}

{Since the expected \ac{isl} is nonnegative and non-increasing, the objective
values converge. The block updates use tight, gradient-consistent
\ac{sca} models~\cite{sun2017mm}. Because the constraints couple the two
blocks, this descent argument alone does not establish joint stationarity or
global optimality.
}
A dense interior-point solution of a convex \ac{qcqp} with $n$ real
decision variables and a comparable number of constraints costs approximately
$\mathcal O(n^{3.5}\log(1/\epsilon))$~\cite{Boyd2004}. Here, the feed and
\ac{rhs} blocks have $2N_cN_{\rm f}K$ and $M$ real decision variables, with
$\mathcal O(N_cK)$ and $\mathcal O(N_cK+M)$ constraints, respectively. {If $C_{\rm v}$ and $C_{\rm m}$ denote the corresponding QCQP solution costs,
the worst-case alternating-SCA cost is
$\mathcal O(I_{\max}R_{\max}B_{\max}(C_{\rm v}+C_{\rm m}))$.
The feasible initialization screens $N_{\rm init}$ \ac{rhs} candidates using
per-subcarrier \ac{zf}, with complexity
$\mathcal O\!\left(
N_{\rm init}N_c(MN_{\rm f}K+N_{\rm f}K^2+K^3)
\right)$, which scales linearly with $N_{\rm init}$.
The proposed method is therefore a centralized, model-based design whose cost
grows with the system dimensions and initialization budget. For larger
networks, distributed or learning-assisted approaches~\cite{dist_askedref} may
reduce online complexity with separate mechanisms to enforce constraints.
}

{
\begin{algorithm}[t]
\caption{\textsc{BlockSCA}: generic block-\ac{sca} update}
\label{alg:block_sca}
\begin{algorithmic}[1]

\Require $\mbf x^{(0)}$, fixed $\mbf y$, $\mathcal P_{\rm x}$,
$\beta_{\rm x}>0$, $\tau_\beta>1$, $R_{\max}$, $B_{\max}$,
$\epsilon_{\rm in}$.

\For{$r=0,\ldots,R_{\max}-1$}\ $\mathtt{accepted}\gets\mathtt{false}$.
    \For{$b=0,\ldots,B_{\max}-1$}\ Solve $\mathcal P_{\rm x}$ at $\mbf x^{(r)}$ for $\mbf x^{+}$.
        \If{$\mbf x^{+}$ is feasible and
        $f_{\rm x}(\mbf x^{+};\mbf y)
        \leq\widehat f_{\rm x}^{(r)}(\mbf x^{+};\mbf y)$} $\ \mathtt{accepted}\gets\mathtt{true}$; \textbf{break}
        \Else $\ \beta_{\rm x}\gets\tau_\beta\beta_{\rm x}$
        \EndIf
    \EndFor
    \If{$\mathtt{accepted}=\mathtt{false}$}\ \Return $(\mbf x^{(r)},\textsc{stalled})$.
    \EndIf\ and\ $\mbf x^{(r+1)}\gets\mbf x^{+}$.
    \If{inner convergence criterion is satisfied}
        \State \Return $(\mbf x^{(r+1)},\textsc{converged})$.
    \EndIf
\EndFor

\State \Return $(\mbf x^{(R_{\max})},\textsc{capped})$.

\end{algorithmic}
\end{algorithm}
}

\vspace{-4mm}

\section{Numerical Results}

\label{sec:numerical_setup}

{We report
${\mathrm{NISL}}
={\bar\Psi_{\rm ISL}}/{\E[|\chi_{0,0}|^2]}$ in dB as
$10\log_{10}({\mathrm{NISL}})$, where
$\E[|\chi_{0,0}|^2]=N_s\sum_n\vartheta_n+N_s^2P_{\rm I}^2$.} \ac{rhs} elements form linearly with spacing $d=\lambda_c/5$, $x_m=(m-(M+1)/2)d$, and 
$[\mbf a_{\rm T}(\theta)]_m=e^{\jj2\pi x_m\sin(\theta)/\lambda_c}/\sqrt M$.
Feeds are uniformly positioned behind aperture at $z_{\rm f}=2\lambda_c$ with their response following the distance-dependent reference-wave model in \cite{rhs_ampcontrol} given by
\begin{equation}
[\widetilde{\mbf G}_n]_{m,r}
\!=\!{e^{-\jj2\pi f_n r_{m,r}/c_0}}/{r_{m,r}},\ 
\mbf G_n
\!=\!{\sqrt M\,\widetilde{\mbf G}_n}
/{\|\widetilde{\mbf G}_n\|_{\rm F}},
\label{eq:G_standard_numerical}
\end{equation}
where $r_{m,r}=\sqrt{(x_m-x_{{\rm f},r})^2+z_{\rm f}^2}$ and
$f_n=f_c+(n-(N_c-1)/2)\Delta f$. We adopt the Rician communication channel as
\begin{multline}
\widetilde{\mbf h}_{n,k}
=
\sqrt{{K_{\rm R}}/({K_{\rm R}+1})}\,
\mbf a_{\rm T}(\phi_{0,k})\\
+\sqrt{{1}/({K_{\rm R}+1})}
\sum_{\ell=1}^{L_c}
\rho_{\ell,k}\mbf a_{\rm T}(\phi_{\ell,k})
e^{-\jj2\pi f_n\tau_{\ell,k}},
\label{eq:rician_channel}
\end{multline}
and $\mbf h_{n,k}=\widetilde{\mbf h}_{n,k}/
\|\widetilde{\mbf h}_{n,k}\|_2$. Here,
$\rho_{\ell,k}\sim\CN(0,1/L_c)$,
$\phi_{0,k},\phi_{\ell,k}\sim\mathcal U[-60^\circ,60^\circ]$, and
$\tau_{\ell,k}\sim\mathcal U[0,100\,\mathrm{ns}]$.

{The target angle is drawn uniformly from $[-30^\circ,30^\circ]$, and normalized QPSK signaling is assumed with $\mu_4=1$ and $\rho=0$. Unless otherwise stated, we set $N_c=8$, $(M,N_{\rm f},K)=(8,2,2)$, $f_c=28$~GHz, $\Delta f=120$~kHz, $K_{\rm R}=10$~dB, $L_c=5$, and $\gamma=6$~dB. The results are averaged over $200$ independent channel/target scenes. We adopt $N_s=16$ as a representative value, since $N_s$ affects neither the feasible set nor the optimization dimensions. Moreover, the scaled expected \ac{isl} takes the form $F_{N_s}=(N_c-1/N_s)A+B$, with $A,B\geq0$, implying $0\leq F_{N_s'}/F_{16}-1\leq(1/16-1/N_s')/(N_c-1/16)$ for any fixed feasible design and $N_s'\geq16$. Thus, base designs remain feasible and tightly bounded for larger $N_s$.
}

The noise variance is fixed as $\sigma_c^2=1$, and the transmit-power budget is expressed through the total transmit \ac{snr} as $P_t=10^{\mathrm{SNR}_{\rm t}/10}$ with $\mathrm{SNR}_{\rm t}=25$~dB. {The illumination requirement is set to $P_0=0.1P_t$. We compare the proposed joint design with four baselines: all-on and random \ac{rhs} configurations whose feed precoders minimize the same expected-\ac{isl} objective, an illumination-maximizing joint design, and a fully digital benchmark minimizing the same objective with $\mbf V_n^{\rm full}\in\C^{M\times K}$ exciting the aperture. 
} For initialization, we test the all-one \ac{rhs} and $N_{\rm init}=10^4$ random vectors from $\mathcal U[0.15,1]$. For each candidate, per-subcarrier \ac{zf} directions are powered to satisfy the constraints, and the feasible candidate with the best objective is selected. We set $\epsilon_{\rm in}=\epsilon_{\rm out}=10^{-3}$, $(R_{\max},I_{\max})=(10,50)$, $\beta_{\rm x}=1$, $B_{\max}=20$, and $\tau_\beta=2$.

\begin{figure}[t]
    \centering
    \includegraphics[width=0.8\columnwidth]{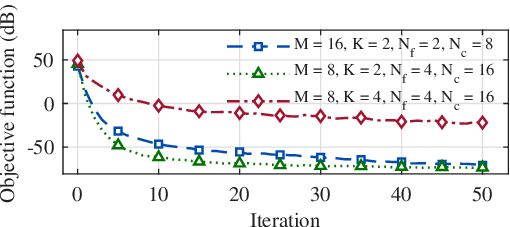}
    \caption{Convergence of the proposed joint \ac{rhs}-\ac{sca} method.}
    \label{fig:convergence}
\vspace{-4mm}
\end{figure}

\begin{table}[t]
\centering

\caption{Median NISL and initialization time versus $N_{\mathrm{init}}$.}
\label{tab:init}
\footnotesize
\setlength{\tabcolsep}{3.0pt}
\begin{tabular}{@{}lccccc@{}}
\toprule
$N_{\mathrm{init}}$ & $10^2$ & $5{\times}10^2$ & $10^3$ & $5{\times}10^3$ & $10^4$ \\
\midrule
Initial NISL (dB) & $-4.2$ & $-5.0$ & $-5.3$ & $-6.1$ & $-6.4$ \\
Final NISL (dB)   & $-144.3$ & $-144.8$ & $-144.9$ & $-145.1$ & $-144.5$ \\
Init.\ time (s)   & 0.13 & 0.61 & 1.22 & 5.97 & 12.0 \\
Init./opt.\ time (\%) & 0.01 & 0.07 & 0.14 & 0.71 & 1.40 \\
\bottomrule
\end{tabular}
\vspace{-4mm}
\end{table}

{Fig.~\ref{fig:convergence} shows a rapid initial decrease followed by gradual
refinement, leading to algorithm convergence.
Table~\ref{tab:init} evaluates $N_{\rm init}$ over 1000 random systems spanning
$M$, $N_c$, $N_{\rm f}$, and $\gamma$. Increasing $N_{\rm init}$ modestly
improves the initial NISL but leaves the final NISL essentially unchanged,
while the initialization time grows approximately linearly. We use
$N_{\rm init}=10^4$ conservatively to reduce initialization sensitivity.
}

\begin{figure}[t]
    \centering
    \includegraphics[width=0.8\columnwidth]{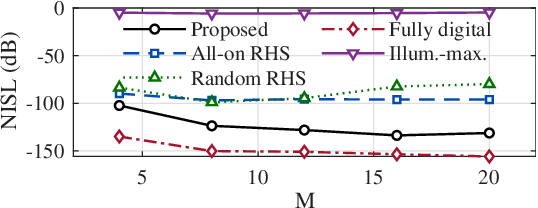}\\[1mm]
    \includegraphics[width=0.8\columnwidth]{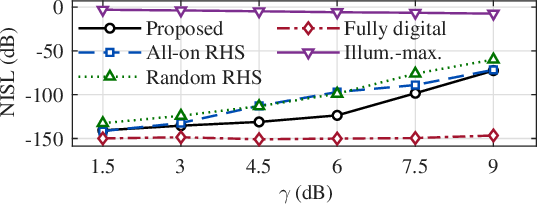}
    \caption{{Average NISL versus (a) $M$ (top) and (b) $\gamma$ (bottom).}}
    \label{fig:main_sweeps}
    \vspace{-3mm}
\end{figure}

{Fig.~\ref{fig:main_sweeps}(a) shows that increasing the number of \ac{rhs}
elements improves the proposed design through greater spatial flexibility,
bringing it closer to the fully digital benchmark.
Fig.~\ref{fig:main_sweeps}(b) shows that stricter \ac{sinr} requirements
progressively reduce the freedom for sidelobe suppression such that the
NISL increases and the proposed design may approach the all-on \ac{rhs} when
the communication constraints become dominant.
}

\begin{figure}[t]
\centering
\includegraphics[width=0.99\columnwidth]{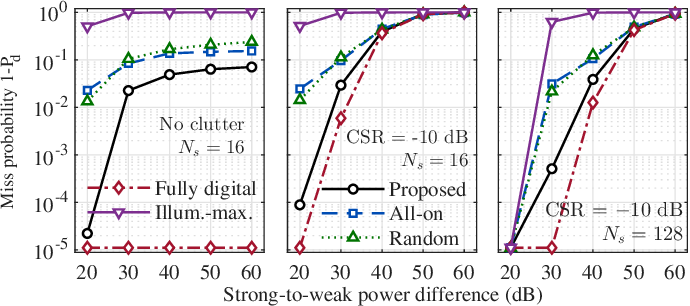}
\caption{{Weak-target miss probability versus the
strong-to-weak target power contrast with $N_c=16$ for different $N_s$ and CSR values. The
receiver noise is set such that strong target has a
post-processing \ac{snr} of $80$~dB at $P_0$.}}
\label{fig:Pdresults}
\vspace{-4mm}
\end{figure}

{Fig.~\ref{fig:Pdresults} evaluates weak-target detection for two
targets at distinct resolvable \ac{rd} cells using two-dimensional
cell-averaging \ac{cfar}~\cite{radar_cfar} with nominal
$P_{\rm fa}=10^{-3}$ over $300$ blocks per scene.
The contrast denotes the strong-to-weak target power difference.
The proposed design consistently improves weak-target detectability
over the \ac{rhs} baselines, particularly at intermediate target
contrasts, while illumination maximization performs poorly, highlighting the importance of
explicit \ac{rd}-sidelobe control.
Stationary clutter from unoptimized directions degrades detection, especially
at higher contrasts, while proposed design keeps its advantage. Nevertheless, $N_s = 128$
substantially improves weak-target detection and extends the
detectable contrast range.
}

\begin{figure}[t]
    \centering
    \includegraphics[width=0.92\columnwidth]{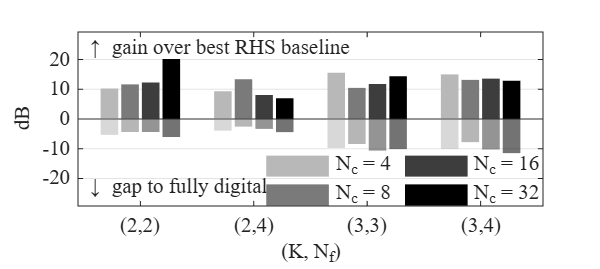}
    \caption{{Scalability gain over the best \ac{rhs} baseline and gap to fully digital.}}
    \label{fig:scale}
\vspace{-5mm}
\end{figure}

{Fig.~\ref{fig:scale} evaluates scalability across different
OFDM grid sizes and user/feed configurations.
The proposed design consistently outperforms the best \ac{rhs}
baseline, confirming the benefits over different setups.
The gains are larger under limited feed resources,
where \ac{rhs} amplitude control provides additional flexibility
for sidelobe suppression. Increasing the feeds also helps narrow the gap
to fully digital beamforming, highlighting the tradeoff between complexity and performance.
}

\vspace{-4mm}
\section{Conclusion and Future Work}

We studied \ac{rd} sidelobe suppression in an \ac{rhs}-assisted \ac{ofdm}-\ac{isac} system. The feed precoders and \ac{rhs} amplitudes were jointly optimized {for the expected \ac{isl} over communication symbols} subject to illumination and communication constraints. We developed an alternating \ac{sca} method to update the variable blocks via convex subproblems while maintaining feasibility. {Numerical results demonstrate the benefit of the joint design across different aperture sizes, communication loads, and \ac{ofdm} dimensions, while showing that the resulting sidelobe suppression can improve target detectability. Future work will consider accelerated and warm-started iterative implementations for faster online adaptation, distributed or learning-assisted solvers~\cite{dist_askedref}, practical RHS hardware models, clutter-aware and multi-direction sensing, and robust designs under sensing and communication uncertainty.}



\vspace{-4mm}

\bibliographystyle{IEEEtran}
\bibliography{ref_abbv}

\end{document}